\documentclass[aps,twocolumn,prl,amsmath,amssymb,floatfix]{revtex4-1}
\usepackage{graphicx}
\makeatletter
\begin{document}
\title{Quantum critical fluctuations,  Planckian dissipation, and compactification scale}

\author{Sudip Chakravarty}
\affiliation{Mani L. Bhaumik Institute for Theoretical Physics\\Department of Physics and Astronomy, University of
California Los Angeles, Los Angeles, California 90095-1547}

\date{\today}
\begin{abstract}
The most striking result here is that the notion of Planckian dissipation is also applicable to the $c$-axis resistivity of the high temperature cuprate superconductors, and to my knowledge this  aspect has not been previously addressed. The derivation involves Kubo formula and  does not require any  mechanism beyond a non-Fermi liquid assumption. The  $c$-axis resistivity, in its essential aspects, is discussed in the context of quantum critical point. Finally, I consider a zero temperature problem with one of its spatial dimensions  compactified.  The warning is that this compatification scale at the quantum critical point behaves similarly to the finite temperature problem, but obviously being at zero temperature there is no  dissipation.
\end{abstract}
\maketitle
 Planckian dissipation has become a widely discussed and important topic~\cite{Sachdev:2005,Zaanen:2004,Zaanen:2019}, not only from the condensed matter point of view but also from  the perspective of high energy physics~\cite{Hartnoll:2018}. Theory also suggests a resolution  in the study of quark-gluon plasma formed at heavy ion colliders~\cite{Policastro:2001}. In cuprates, to my knowledge, all such discussion has involved in-plane resistivity. It is quite remarkable that when properly interpreted the notion is also valid for the $c$-axis (perpendicular to the plane) resistivity.  The purpose of this paper is to  elucidate this aspect and accomplish it by examining the notion of Planckian dissipation and its connection to quantum criticality~\cite{Chakravarty:1988,Chakravarty:1989}.

Consider the conductivity sum rule~\cite{Kubo:1957}
applied to cuprate superconductors.  The sum rule applies independently to  $ab$-plane and to $c$-direction, perpendicular to the CuO-planes. The $ab$-plane conductivity was invoked previously~\cite{Hirsch:1992}; here we consider the $c$-axis sum rule~\cite{Chakravarty:1999} which is simpler to apply  by comparison. One need not solve a strongly interacting correlated electron problem.  Consider the 
full 
hamiltonian 
$H=H_{\rm rest}+H_c$; the $c$-axis kinetic energy is defined by 
\begin{equation} 
H_c = -\sum_{ij,s}t_{\perp}(ij,l) c^{\dagger}_{il,s}c_{jl+1,s}+{\rm h. 
c.}
\end{equation} 
The remainder, $H_{\rm rest}$, contains no further inter-plane interaction 
terms, but it is 
otherwise arbitrary and may contain impurity interactions  that couple 
to the charge density. 
The hopping matrix element $t_{\perp}(ij,l)$, where  $(i,j)$ refers to 
the sites of 
the two-dimensional 
lattice, and  $l$ to the layer index. The hopping matrix element  can be random in the presence of 
impurities . The 
electron operators 
$(c,c^{\dagger})$ are 
also labeled by a spin index 
$s$. We denote the magnitude of $t_{\perp}(ij,l)$ by $t_{\perp}$. 
One can  adapt a sum rule derived first by  Kubo \cite{Kubo:1957} 
to get a sum rule 
for the $c$-axis 
optical conductivity 
$\sigma^c(\omega,T)$, which is 
\begin{equation} 
\int_{0}^{\omega_{c}}d\omega\ {\rm Re}\ \sigma^c(\omega,T)={\frac{\pi e^2 d^2}
{2\hbar^2 
(Ad)}}\langle -H_c\rangle. 
\label{srule} 
\end{equation} 
Here the 
average refers to the quantum statistical average,  $A$ to 
the two-dimensional 
area, and 
$d$ 
 to the separation 
between 
the CuO-planes of the unit cells. 
The hamiltonian $H_c$  is an effective hamiltonian valid only for low 
energy 
processes, $\omega<\omega_{c}$ that do  not involve inter-band 
transitions, and cannot exceed of order $2-3$ eV.  
 
Since 
$H_c$ is a low-energy effective hamiltonian, the upper 
limit in Eq.~(\ref{srule})  cannot exceed 
an inter-band cutoff $\omega_c$, of order of a few electron volts. Beyond this we need not be more specific, because our goal is to 
deduce results as $T\to 0$. 

To estimate the right hand side, one must make a distinction between Fermi and non-Fermi liquids. While there 
is no universal theory of non-Fermi liquids, its characterization in terms of spectral function is very useful.We would like to build a model in which the non-Fermi liquid behavior is 
 essential, which we briefly recall: the  analytic
continuation of the Green's function to the second Riemann sheet contains branch points instead
of simple poles. A spectral function $A$  that satisfies the
scaling relation
\begin{equation}
A(\Lambda^{y_1}[k-k_F],\Lambda^{y_2}\omega)= \Lambda^{y_A}A([k-k_F],\omega),
\end{equation}
where $y_1$, $y_2$, and $y_A$ are the  exponents defining the universality class of
the critical Fermi system. The values of the exponents other than the set
$y_1=1$, $y_2=1$, and $y_A=-1$ 
represent a non-Fermi liquid.  

For a  non-Fermi liquid, an electron creation operator of wavevector {\bf k} and spin $\uparrow$ acting
on the ground state  creates a linear {\em superposition of states} that carry the momentum $\bf k$ and spin $\uparrow$. 
However, the act of
inserting an electron into a non-Fermi liquid cannot be renormalized away by defining a single
quasiparticle:  but  the excitation energy is not uniquely related to $\bf k$. The  orthogonality catastrophe 
generally leads to ground state overlaps of the form
\begin{equation}
\langle N| c_{{\bf k}\sigma}|N+1\rangle\approx
e^{-\alpha\int_{\omega_0({\bf k})}^{\omega_c}{d\omega \over \omega}}=\left({\omega_0({\bf k})\over
\omega_c}\right)^{\alpha} .\label{ortho}
\end{equation}
The quantity $\omega_0$ is a low energy cutoff and $\alpha$ is an orthogonality exponent that will depend on electron-electron interactions. Clearly the overlap vanishes as $\omega_0\to 0$. 
The above expression is simply the $z$-factor which must vanish at the Fermi surface in a non-Fermi
liquid state.  In the superconducting state, the states far above the gap are similar to a non-Fermi liquid  normal
state. However, in the regions of the Brillouin zone where there is a 
gap, the orthogonality catastrophe will be cut off because $\omega_0$ in
Eq.~(\ref{ortho}) will be of the order of the energy gap $\Delta$, and the overlap factor will be $({\Delta\over \omega_c})^{\alpha}$. In the regions where the gap is vanishingly small, orthogonality
catastrophe will act with full force.  

Thus, the average $\langle H_c \rangle$  in the exact ground state $|\widehat{0}\rangle$  to lowest order in $t_{\perp}^2/W$ 
in a perturbative expansion is
\begin{equation} 
\langle \widehat{0}|H_c|\widehat{0}\rangle=-2\sum_{n\ne 0}\frac{|\langle 
0|H_c|n\rangle|^2}{E_n-E_{0}}+\cdots, 
\label{secondorder}
\end{equation} 
where $E_n$ and $|n\rangle$ on the right are the eigenvalues and the eigenstates of 
$H_{\rm rest}$. The first order term is zero, because of the absence  
of diagonal coherent single particle tunneling between the unit cells,  which is due to non-fermi liquid nature of the in-plane excitations.

For conserved parallel momentum, the expansion on the right hand side of 
Eq.~(\ref{secondorder}) would lead to  
vanishing energy denominators invalidating the expansion. 
In a non-Fermi liquid state, however, the matrix 
elements should vanish  for 
vanishing energy differences, and  the the sum is likely to be skewed to 
high energies.  Thus,  the energy 
denominator can be approximated by $W$, of the order of the bandwidth, and the sum can 
be collapsed using the completeness 
condition to 
$\langle 0|H_c^2|0\rangle/W$. 
Referring to Eq.~\ref{ortho} and Eq.~\ref{secondorder}, the difference between 
$\int_0^{W}d\omega\omega^{2\alpha}/\omega$ and 
$\int_0^{W}d\omega\omega^{2\alpha}/W$ is an unimportant 
numerical factor that can be absorbed in 
the definition of  $W$. 
Thus the effective hamiltonian is $-H_c^2/W$. 

On dimensional grounds $c$-axis conductivity is 
\begin{equation} 
\sigma_c(T)=a\left(\frac{e^2 d t_{\perp}^2}{ A W \hbar^2}\right){\frac{1}
{\Omega(T)}}, 
\label{conductivity} 
\end{equation} 
where $a$ is a numerical constant weakly dependent on the band 
structure. The inelastic 
scattering time  is proportional to the unknown function $1/\Omega(T)$. 

We apply the $c$-axis sum rule at two different temperatures $T_{1}$ and $T_{2}$. Quite generally in the 
superconducting state, $\sigma^{cs}(\omega,T_{1})=D_c(T_{1})\delta(\omega)+ 
\sigma^{cs}_{\rm reg}(\omega,T_{1})$, where $D_c(T_{1})$ is the superfluid 
weight. From 
Eq.~(\ref{srule}), it follows that 
\begin{eqnarray} 
D_c(T_1)&=&\int_{0^+}^{\omega_c}d\omega \bigg[{\rm Re}\ 
\sigma^{cn}(\omega,T_2)-{\rm Re}\ \sigma^{cs}_{\rm 
reg}(\omega,T_1)\bigg]\nonumber\\ 
&+&{\pi e^2 d^2\over 2 Ad \hbar^2}\bigg[\langle 
-H_c(T_1)\rangle_s-\langle 
-H_c(T_2)\rangle_n\bigg]. 
\label{srule2} 
\end{eqnarray} 
Here, if $T_2 < T_c$, $\langle -H_c(T_2)\rangle_n$ is to be understood 
as taken in 
the normal state extrapolated  below $T_c$, and $\sigma^{cn}$ is the 
corresponding conductivity. Let
$T_1=T_2=0$ in Eq.~(\ref{srule2}). 
From the experiments of Ando {\em et al.\/}\cite{Ando}, it is seen  that 
the $c$-axis response in the 
normal state obtained by destroying superconductivity is insulating as 
$T\to 0$; it follows  that 
$\sigma^{cn}(\omega, T=0)\sim \omega^2$, as $\omega\to 0$. The regular part 
$\sigma^{cs}_{\rm reg}(\omega, T=0)$ is also expected to vanish as a 
power law . At high frequencies  the two conductivities 
must, however, approach each other. Consequently it is reasonable to 
hypothesize that the conductivity integral on the right hand side of 
Eq.~(\ref{srule2}) is negligibly small. Therefore, 
\begin{equation} 
D_c(0) \equiv {c^2 \over 8\lambda_c^2} \approx {\pi e^2 d^2\over 2 Ad 
\hbar^2}\bigg[\langle 
-H_c \rangle_s-\langle -H_c \rangle_n\bigg], 
\label{energy} 
\end{equation} 
where we assumed local 
London electrodynamics.  In any case, the right hand side 
should be a lower bound. 
Then 
\begin{equation} 
\frac{c^2}{8 \lambda_c^2}=\frac{4\pi}{a}  \sigma_c(T) \Omega(T) 
\left[u_s-u_n\right], 
\label{penetration} 
\end{equation} 
where $u_{s,n}$ is $\langle 
0|(H_c/t_{\perp})^2|0\rangle_{s,n}$. The average here is 
with respect to the ground state of $H_{\rm rest}$. The subscripts $s$ and $n$ refer to the superconducting and the normal ground states. By normal  state I mean a state in which superconductivity is destroyed. Presently it can be achieved reasonably well by applying high magnetic field~\cite{Taillefer:2007}. Only temperature dependencies  are contained in the product $\sigma_c(T) \Omega(T)$.

\begin{figure}[hbtp]
\includegraphics[width=\linewidth]{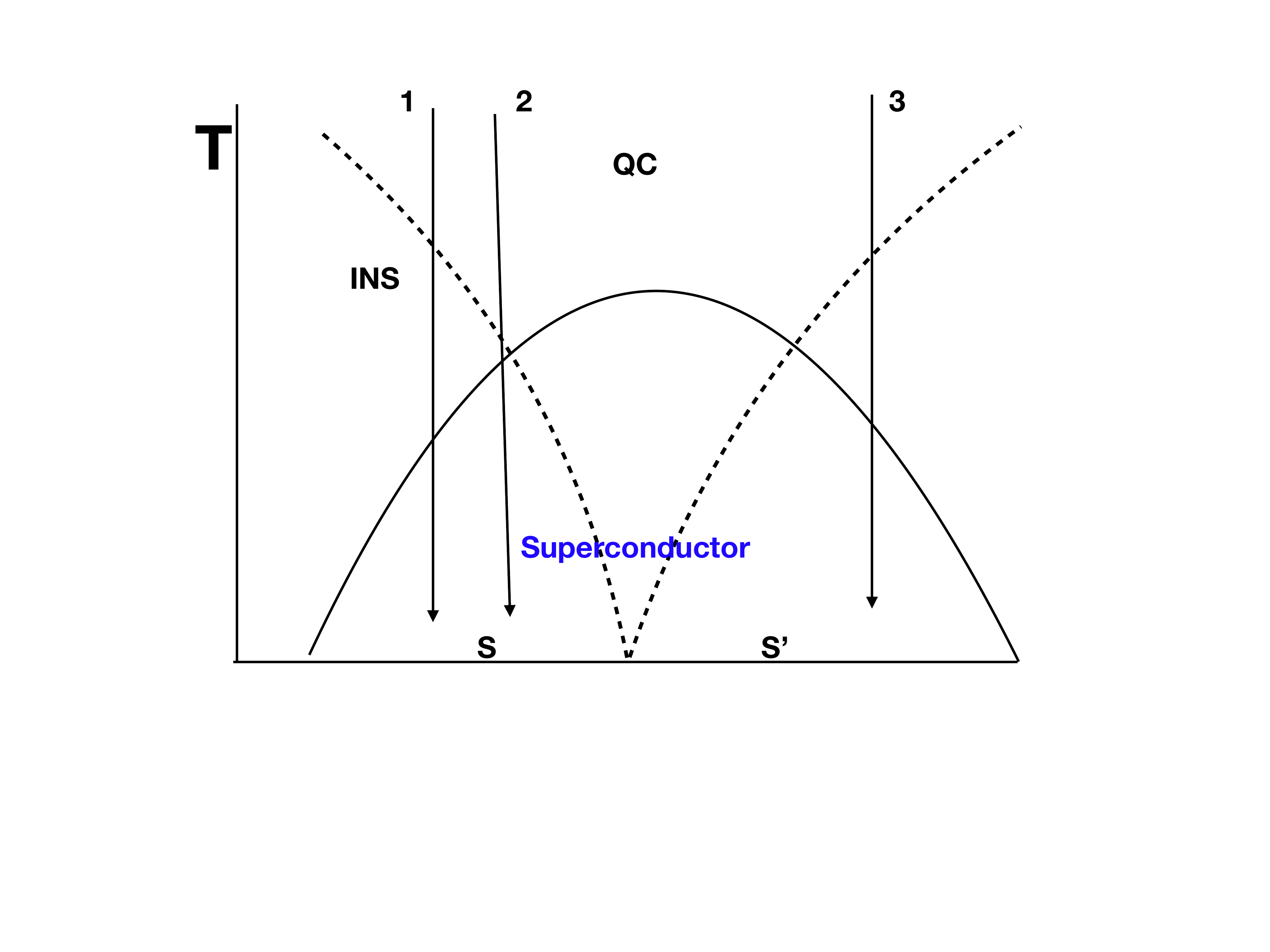}
\caption{Dimensional crossover of $\rho_{c}$ as a function of temperature $T$. INS stands for insulator and QC for quantum critical. Here S and S' are two phases, both with superconducting order parameter. For example, S may be $d$-density wave along with superconductivity and S' without the $d$-density wave.  Thus there can be a continuous phase transition. Three vertical arrows are possible experimental trajectories that intersect with quantum critical fan describing a crossover scale~\cite{Chakravarty:1988,Chakravarty:1989}. These trajectories cross over to the superconducting state. {\em There is no reason for the quantum critical point to be in the middle of the superconducting dome as drawn here for simplicity.} The line 2 is shown to have  very little insulating up turn.}
\label{phase}
\end{figure}

To make progress, we note that the $c$-axis resistivity, 
$\rho_c(T)$, can often be fitted  ($p>0$) to a combination of power laws. This power law behavior is not necessary as long as there is a quantum critical crossover. The present form is chosen for convenience: 
\begin{equation} 
\rho_c(T)=b_1 T^{-p}+b_2'T. 
\end{equation} 
In some materials such as $\mathrm{YBa_2Cu_{3}O_{6+\delta}}$ the $c$-axis resistivity may not be substantially larger 
from the $ab$-plane resistivity close to $T_{c}$, and may not have a substantial insulating  up turn. From the picture, see line 2 in Fig.~\ref{phase}, it is easy to understand  to be the case when  the the experimental trajectory is very close to the quantum critical line and the superconducting boundary.

We express 
Eq.~(\ref{penetration}) in terms of the temperature $T^*$ (not the notation for the pseudogap) at which the 
c-axis resistivity takes its minimum value given the empirical behavior of the $c$-axis resistivity.
We get 
\begin{equation} 
\frac{1}{\lambda_c^2}=\sigma_c(T^*) T^* 
\left\{\frac{4\pi}{c^2} \frac{b_2(p+1)}{ p}\left[u_s-u_n\right]\right\}, 
\label{Eq8}
\end{equation} 
where $b_2=b_2'(d e^2 t_{\perp}^2/\hbar^2 A W)$. 
The expression in the curly brackets depends dominantly on $b_2$, 
which describes the high temperature linear
resistivity. The low temperature behavior enters only through the 
exponent 
$p$, but of course cuts off at $T_c$. What could be the meaning of $T^*$? At a trivial level it is a
lower bound to $\rho_c$. I conjecture that it is also the boundary of the quantum critical region above which the linear behavior 
of the resistivity appears.
\begin{figure}[htbp]
\includegraphics[width=\linewidth]{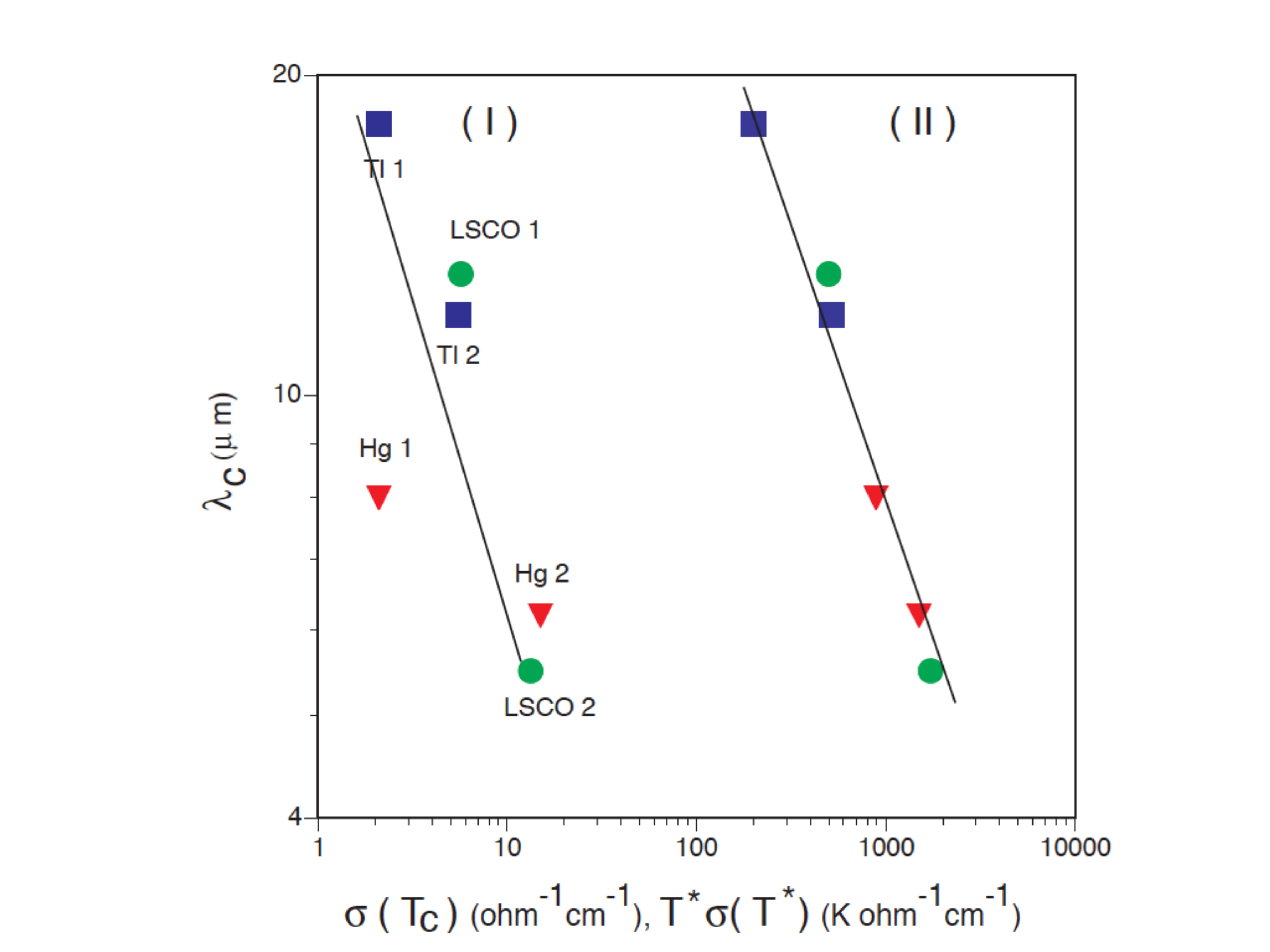}
\caption{The Basov plot: $\ln \lambda_c$ is plotted against  $\ln 
\sigma_c(T_c)$, as in the original 
Basov plot (I), and against $\ln [T^*\sigma_c(T^*)]$ as discussed here 
(II). The legends in group (II) are the same as those in group (I). Tl1: Ref.~\cite{Moler:1998} Tl2: Ref.~\cite{Basov:1999}; 
Hg1: Ref.~\cite{Kirtley:1998}, Hg2: Ref.~\cite{Basov:1999}; LSCO 1 (12\%), LSCO 2 (15\%): Refs.~\cite{Loram:1994,
Uchida:1996,Nakamura:1993}.}
\label{Fig1}
\end{figure}

Thus, provided the expression in curly brackets is a universal 
constant, a plot of $\ln \lambda_c$ against 
$\ln[\sigma_c(T^*) T^*]$ should be a universal straight line, 
independent of material, and temperature $T^{*}$, which is {\em indeed} the case, thus validating Eq.~(\ref{Eq8}).
Basov {\em 
et al.}\cite{Basov:1994} suggested a similar correlation by plotting 
$\ln \lambda_c$ against 
$\sigma_c(T_c)$, shown as (I) in Fig.~\ref{Fig1} (reproduced from~\cite{Chakravarty:1999}). In comparison to Basov correlation, the correlation 
discussed here, shown as (II), is excellent. 
In Fig.~\ref{Fig1}, we have taken $T^*\approx T_c$ 
for those  doped 
materials that show simply a flattening of $\rho_{c}(T)$ close to $T_c$; see the discussion above in reference to Fig.~\ref{phase}.  It is at least consistent therefore to assume on dimensional grounds that $\sigma(T^{*})\propto \hbar/k_{B} T^{*}$, indeed a signature of Planckian dissipation. Finally, if $T^{*}$ is any temperature $T$, which lies within the quantum critical fan, a stronger result $\sigma(T)\propto \hbar/k_{B} T$ is plausible. This follows from the fact that the properties in the fan are dictated by the $T=0$ quantum critical point barring crossover scales. From the Basov plot, such a strong plausibility argument is not possible. The extra factor of temperature in the product $\sigma(T) T$, as theoretically predicted,  is {\em essential}, and was empirically rediscovered by Homes {\em et al}~\cite{Homes:2004} with $T$ replaced arbitrarily by $T_{c}$. 

To substantiate the above remark (replacing $T^{*}$ by $T$) note that the correlation length is largely given by the quantum critical fan~\cite{Chakravarty:1989} within  which the physics is largely dictated by the QCP until crossovers take place. A simple one-loop example of crossover is shown below. 
The full expression for the correlation length  in one loop approximation~\cite{Chakravarty:1989} is,

\begin{equation}
\frac{1}{\xi} = \sqrt{\frac{2}{\pi}}\left(\frac{k_{B}T}{\hbar c}\right) \sinh^{-1}\left(\frac{e}{2}e^{-2\pi\rho_{s}/k_{B}T}\right).
\end{equation}
Close to the quantum critical point ($\rho_{s}\to 0$),
\begin{equation}
\frac{1}{\xi}\approx 0.9 \left(\frac{k_{B}T}{\hbar c}\right)  (1-4.5 \rho_{s}/k_{B}T) ,
\end{equation}
Clearly $\xi$  is a result of fluctuations at the quantum critical point. The length scale at the QCP does not reflect  a free particle or even a single quasiparticle, but rather a scale invariant description of excitations made {\em explicit} by the Euclidean field theory.

Thus, 
we see that $b_2$ in Eq.~\ref{Eq8} is indeed inversely proportional to 
$[u_s-u_n]$, which is proportional to the $T=0$ superfluid 
density, 
$n_s^c(0)$.The inverse proportionality of $b_{2}$ with $\left[u_s-u_n\right]$ can be tested in experiments, but has not been thus far. 
\begin{figure}[hbtp]
\includegraphics[scale=0.3]{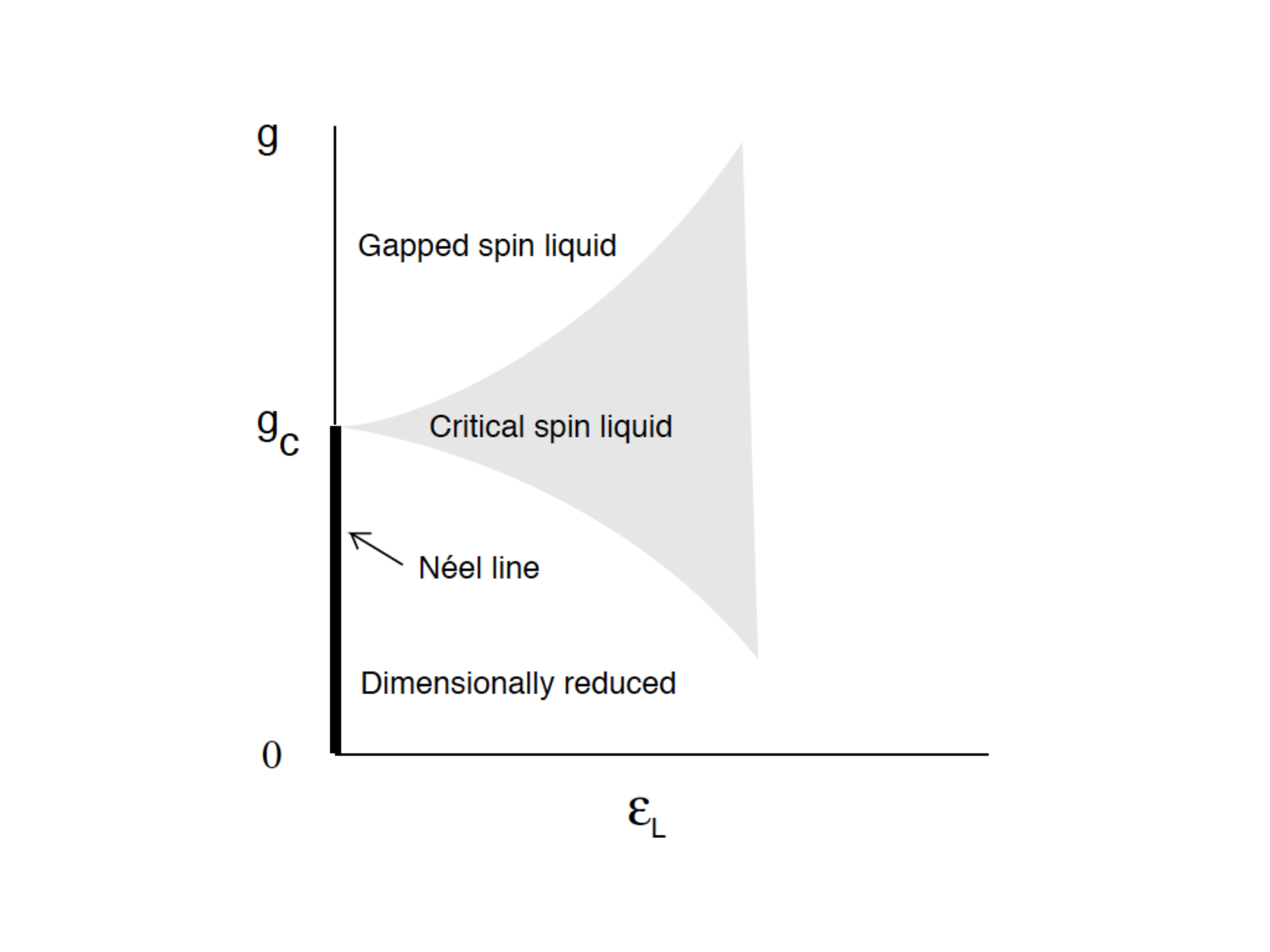}
\caption{Dimensional crossover at $T=0$ in $d=2$ Here  $g$ is a dimensionless coupling constant and $\varepsilon = \frac{\hbar c}{L}$ is an energy-like variable arising from the compactification of $L_{x}$.\label{Fig2}}
\end{figure}

We now turn to a simple example~\cite{Chakravarty:1996} where such a connection to dissipation cannot be made, but in all other respects the problem is similar to the $O(3)$ non-linear $\sigma$-model, where  $\hat{\Omega}$ is a three component unit vector, at finite temperature, at least insofar as the quantum critical fan is concerned. 
Consider   a finite $L_{x}$ giving a compactification length, but $\beta \hbar = \infty$. This model has no dissipation but behaves in an analogous manner.  \begin{equation}
\frac{S}{\hbar}=\frac{\rho_s^0}{ 2\hbar}\int_0^{\beta\hbar}d\tau\int
d^{d-1}x
\int_0^L
dx_1\left[(\partial_{\mu}\hat\Omega)^2+\frac{1}
{c^2}\left(\frac{\partial\hat\Omega}{\partial\tau}\right)^2\right],
\end{equation}

Lengths in all directions,
except $x_1$, are infinite;   $x_1$ is compactified by a  periodic
boundary condition. The  problem at $T=0$ is 
isomorphic to the problem at finite ``temperature", where the
temperature-like variable is
$\varepsilon_L=\hbar c/L_{x}$, with the  identifications of the
dimensionless parameters,
$g_0=\frac{\hbar c \Lambda^{d-1}}{ \rho_s^0}$ and
$\varepsilon^0_L=\frac{\hbar c\Lambda^{d-2}}{ L_{x}\rho_s^0}$,
where $\Lambda$ is an arbitrary high energy cutoff, the energy like parameter $\varepsilon^0_L$ plays the role of the
dimensionless temperature-like variable. The relevant excitation velocity will remain unspecified.

There is a quantum critical point with scaie invariant quantum fluctuations, but no dissipation is involved in this problem, merely quantum fluctuations at the quantum critical point. In Fig.~\ref{Fig2} (reproduced from~\cite{Chakravarty:1996}), the N\'eel line is separated from a gapped spin liquid with an intervening region labeled as critical spin liquid. Originally, this model was used to predict a crossover and deduce the spin gaps in spin ladders, which agreed quite remarkably with numerical calculations. 

Later the model was extended~\cite{Chakravarty:1997} to finite temperatures to deduce successive crossovers in temperatue, which can be analyzed in terms of Planckian  dissipation scales that we leave for future investigations. 
But to repeat, the $T=0$ model has nothing to do with  dissipation. 

Recall that in order to discuss Planckian dissipation above we had to introduce a sum rule due to  Kubo  at $T\ne 0$. This sum rule applies independently for any direction. The  $c$-axis sum rule has not been discussed much in this context; see, however, \cite{Chakravarty:1999}. It is notable that I could combine a simple perturbative expansion, thanks to non-Fermi liquid assumption,  with some ``empirical'' results concerning the $c$-axis resistivity, which is physically clearly justified  in Fig.~\ref{phase}. Of course an appeal to universality is implicit. I could have used the $ab$-plane sum rule but in that case a perturbative analysis would not have been applicable because of strong interactions.  Later we relied on quantum criticality to arrive at the dissipation scale. Is there any interpretation that we can give to  the $T=0$ problem above?  A rigorous answer does not exist.   I finish by emphasizing that the existence of quantum criticality does not always lead to time scales relevant for  dissipation, as discussed in the counterexample.  Perhaps one can extend much of the theory to the criticality in the overdoped regime of cuprates~\cite{Kopp:2007}, recently confirmed in Refs.~\cite{Kurashima:2018} and~\cite{Sarkar:2020}.

This work was performed at the Aspen Center for Physics, which is supported by National Science Foundation grant PHY-1607611.
It  was also partially supported by funds from David S. Saxon Presidential Term Chair at the University of California Los Angeles. I thank M. Mulligan. S. Kivelson and S. Raghu for help  and Chandra Varma for general comments.


\begin{thebibliography}{24}%
\makeatletter
\providecommand \@ifxundefined [1]{%
 \@ifx{#1\undefined}
}%
\providecommand \@ifnum [1]{%
 \ifnum #1\expandafter \@firstoftwo
 \else \expandafter \@secondoftwo
 \fi
}%
\providecommand \@ifx [1]{%
 \ifx #1\expandafter \@firstoftwo
 \else \expandafter \@secondoftwo
 \fi
}%
\providecommand \natexlab [1]{#1}%
\providecommand \enquote  [1]{``#1''}%
\providecommand \bibnamefont  [1]{#1}%
\providecommand \bibfnamefont [1]{#1}%
\providecommand \citenamefont [1]{#1}%
\providecommand \href@noop [0]{\@secondoftwo}%
\providecommand \href [0]{\begingroup \@sanitize@url \@href}%
\providecommand \@href[1]{\@@startlink{#1}\@@href}%
\providecommand \@@href[1]{\endgroup#1\@@endlink}%
\providecommand \@sanitize@url [0]{\catcode `\\12\catcode `\$12\catcode
  `\&12\catcode `\#12\catcode `\^12\catcode `\_12\catcode `\%12\relax}%
\providecommand \@@startlink[1]{}%
\providecommand \@@endlink[0]{}%
\providecommand \url  [0]{\begingroup\@sanitize@url \@url }%
\providecommand \@url [1]{\endgroup\@href {#1}{\urlprefix }}%
\providecommand \urlprefix  [0]{URL }%
\providecommand \Eprint [0]{\href }%
\providecommand \doibase [0]{http://dx.doi.org/}%
\providecommand \selectlanguage [0]{\@gobble}%
\providecommand \bibinfo  [0]{\@secondoftwo}%
\providecommand \bibfield  [0]{\@secondoftwo}%
\providecommand \translation [1]{[#1]}%
\providecommand \BibitemOpen [0]{}%
\providecommand \bibitemStop [0]{}%
\providecommand \bibitemNoStop [0]{.\EOS\space}%
\providecommand \EOS [0]{\spacefactor3000\relax}%
\providecommand \BibitemShut  [1]{\csname bibitem#1\endcsname}%
\let\auto@bib@innerbib\@empty
\bibitem [{\citenamefont {Sachdev}(2005)}]{Sachdev:2005}%
  \BibitemOpen
  \bibfield  {author} {\bibinfo {author} {\bibfnamefont {S.}~\bibnamefont
  {Sachdev}},\ }\href@noop {} {\emph {\bibinfo {title} {Quantum phase
  transitions}}}\ (\bibinfo  {publisher} {Cambridge University Press},\
  \bibinfo {year} {2005})\BibitemShut {NoStop}%
\bibitem [{\citenamefont {Zaanen}(2004)}]{Zaanen:2004}%
  \BibitemOpen
  \bibfield  {author} {\bibinfo {author} {\bibfnamefont {J.}~\bibnamefont
  {Zaanen}},\ }\href {\doibase 10.1038/430512a} {\bibfield  {journal} {\bibinfo
   {journal} {Nature}\ }\textbf {\bibinfo {volume} {430}},\ \bibinfo {pages}
  {512} (\bibinfo {year} {2004})}\BibitemShut {NoStop}%
\bibitem [{\citenamefont {Zaanen}(2019)}]{Zaanen:2019}%
  \BibitemOpen
  \bibfield  {author} {\bibinfo {author} {\bibfnamefont {J.}~\bibnamefont
  {Zaanen}},\ }\href {\doibase 10.21468/SciPostPhys.6.5.061} {\bibfield
  {journal} {\bibinfo  {journal} {SciPost Phys.}\ }\textbf {\bibinfo {volume}
  {6}},\ \bibinfo {pages} {61} (\bibinfo {year} {2019})}\BibitemShut {NoStop}%
\bibitem [{\citenamefont {Hartnoll}\ \emph {et~al.}()\citenamefont {Hartnoll},
  \citenamefont {Lucas},\ and\ \citenamefont {Sachdev}}]{Hartnoll:2018}%
  \BibitemOpen
  \bibfield  {author} {\bibinfo {author} {\bibfnamefont {S.~A.}\ \bibnamefont
  {Hartnoll}}, \bibinfo {author} {\bibfnamefont {A.}~\bibnamefont {Lucas}}, \
  and\ \bibinfo {author} {\bibfnamefont {S.}~\bibnamefont {Sachdev}},\
  }\href@noop {} {\emph {\bibinfo {title} {Holographic Quantum Matter}}}\
  (\bibinfo  {publisher} {MIT Press, 2018})\BibitemShut {NoStop}%
\bibitem [{\citenamefont {Policastro}\ \emph {et~al.}(2001)\citenamefont
  {Policastro}, \citenamefont {Son},\ and\ \citenamefont
  {Startinets}}]{Policastro:2001}%
  \BibitemOpen
  \bibfield  {author} {\bibinfo {author} {\bibfnamefont {G.}~\bibnamefont
  {Policastro}}, \bibinfo {author} {\bibfnamefont {D.~T.}\ \bibnamefont {Son}},
  \ and\ \bibinfo {author} {\bibfnamefont {A.~O.}\ \bibnamefont {Startinets}},\
  }\href@noop {} {\bibfield  {journal} {\bibinfo  {journal} {Phys. Rev. Lett.}\
  }\textbf {\bibinfo {volume} {87}},\ \bibinfo {pages} {081601} (\bibinfo
  {year} {2001})}\BibitemShut {NoStop}%
\bibitem [{\citenamefont {Chakravarty}\ \emph {et~al.}(1988)\citenamefont
  {Chakravarty}, \citenamefont {Halperin},\ and\ \citenamefont
  {Nelson}}]{Chakravarty:1988}%
  \BibitemOpen
  \bibfield  {author} {\bibinfo {author} {\bibfnamefont {S.}~\bibnamefont
  {Chakravarty}}, \bibinfo {author} {\bibfnamefont {B.~I.}\ \bibnamefont
  {Halperin}}, \ and\ \bibinfo {author} {\bibfnamefont {D.~R.}\ \bibnamefont
  {Nelson}},\ }\href {\doibase 10.1103/PhysRevLett.60.1057} {\bibfield
  {journal} {\bibinfo  {journal} {Phys. Rev. Lett.}\ }\textbf {\bibinfo
  {volume} {60}},\ \bibinfo {pages} {1057} (\bibinfo {year}
  {1988})}\BibitemShut {NoStop}%
\bibitem [{\citenamefont {Chakravarty}\ \emph {et~al.}(1989)\citenamefont
  {Chakravarty}, \citenamefont {Halperin},\ and\ \citenamefont
  {Nelson}}]{Chakravarty:1989}%
  \BibitemOpen
  \bibfield  {author} {\bibinfo {author} {\bibfnamefont {S.}~\bibnamefont
  {Chakravarty}}, \bibinfo {author} {\bibfnamefont {B.~I.}\ \bibnamefont
  {Halperin}}, \ and\ \bibinfo {author} {\bibfnamefont {D.~R.}\ \bibnamefont
  {Nelson}},\ }\href {\doibase 10.1103/PhysRevB.39.2344} {\bibfield  {journal}
  {\bibinfo  {journal} {Phys. Rev. B}\ }\textbf {\bibinfo {volume} {39}},\
  \bibinfo {pages} {2344} (\bibinfo {year} {1989})}\BibitemShut {NoStop}%
\bibitem [{\citenamefont {Kubo}(1957)}]{Kubo:1957}%
  \BibitemOpen
  \bibfield  {author} {\bibinfo {author} {\bibfnamefont {R.}~\bibnamefont
  {Kubo}},\ }\href {\doibase 10.1143/JPSJ.12.570} {\bibfield  {journal}
  {\bibinfo  {journal} {Journal of the Physical Society of Japan}\ }\textbf
  {\bibinfo {volume} {12}},\ \bibinfo {pages} {570} (\bibinfo {year}
  {1957})}\BibitemShut {NoStop}%
\bibitem [{\citenamefont {Hirsch}(1992)}]{Hirsch:1992}%
  \BibitemOpen
  \bibfield  {author} {\bibinfo {author} {\bibfnamefont {J.~E.}\ \bibnamefont
  {Hirsch}},\ }\href {\doibase https://doi.org/10.1016/0921-4534(92)90415-9}
  {\bibfield  {journal} {\bibinfo  {journal} {Physica C: Superconductivity}\
  }\textbf {\bibinfo {volume} {199}},\ \bibinfo {pages} {305 } (\bibinfo {year}
  {1992})}\BibitemShut {NoStop}%
\bibitem [{\citenamefont {Chakravarty}\ \emph {et~al.}(1999)\citenamefont
  {Chakravarty}, \citenamefont {Kee},\ and\ \citenamefont
  {Abrahams}}]{Chakravarty:1999}%
  \BibitemOpen
  \bibfield  {author} {\bibinfo {author} {\bibfnamefont {S.}~\bibnamefont
  {Chakravarty}}, \bibinfo {author} {\bibfnamefont {H.-Y.}\ \bibnamefont
  {Kee}}, \ and\ \bibinfo {author} {\bibfnamefont {E.}~\bibnamefont
  {Abrahams}},\ }\href {\doibase 10.1103/PhysRevLett.82.2366} {\bibfield
  {journal} {\bibinfo  {journal} {Phys. Rev. Lett.}\ }\textbf {\bibinfo
  {volume} {82}},\ \bibinfo {pages} {2366} (\bibinfo {year}
  {1999})}\BibitemShut {NoStop}%
\bibitem [{\citenamefont {Doiron-Leyraud}\ \emph {et~al.}(2007)\citenamefont
  {Doiron-Leyraud}, \citenamefont {Proust}, \citenamefont {LeBoeuf},
  \citenamefont {Levallois}, \citenamefont {Bonnemaison}, \citenamefont
  {Liang}, \citenamefont {Bonn}, \citenamefont {Hardy},\ and\ \citenamefont
  {Taillefer}}]{Taillefer:2007}%
  \BibitemOpen
  \bibfield  {author} {\bibinfo {author} {\bibfnamefont {N.}~\bibnamefont
  {Doiron-Leyraud}}, \bibinfo {author} {\bibfnamefont {C.}~\bibnamefont
  {Proust}}, \bibinfo {author} {\bibfnamefont {D.}~\bibnamefont {LeBoeuf}},
  \bibinfo {author} {\bibfnamefont {J.}~\bibnamefont {Levallois}}, \bibinfo
  {author} {\bibfnamefont {J.-B.}\ \bibnamefont {Bonnemaison}}, \bibinfo
  {author} {\bibfnamefont {R.}~\bibnamefont {Liang}}, \bibinfo {author}
  {\bibfnamefont {D.~A.}\ \bibnamefont {Bonn}}, \bibinfo {author}
  {\bibfnamefont {W.~N.}\ \bibnamefont {Hardy}}, \ and\ \bibinfo {author}
  {\bibfnamefont {L.}~\bibnamefont {Taillefer}},\ }\href@noop {} {\bibfield
  {journal} {\bibinfo  {journal} {Nature}\ }\textbf {\bibinfo {volume} {447}},\
  \bibinfo {pages} {565} (\bibinfo {year} {2007})}\BibitemShut {NoStop}%
\bibitem [{\citenamefont {Moler}\ \emph {et~al.}(1998)\citenamefont {Moler},
  \citenamefont {Kirtley}, \citenamefont {Hinks}, \citenamefont {Li},\ and\
  \citenamefont {Xu}}]{Moler:1998}%
  \BibitemOpen
  \bibfield  {author} {\bibinfo {author} {\bibfnamefont {K.~A.}\ \bibnamefont
  {Moler}}, \bibinfo {author} {\bibfnamefont {J.~R.}\ \bibnamefont {Kirtley}},
  \bibinfo {author} {\bibfnamefont {D.~G.}\ \bibnamefont {Hinks}}, \bibinfo
  {author} {\bibfnamefont {T.~W.}\ \bibnamefont {Li}}, \ and\ \bibinfo {author}
  {\bibfnamefont {M.}~\bibnamefont {Xu}},\ }\href {\doibase
  10.1126/science.279.5354.1193} {\bibfield  {journal} {\bibinfo  {journal}
  {Science}\ }\textbf {\bibinfo {volume} {279}},\ \bibinfo {pages} {1193}
  (\bibinfo {year} {1998})}\BibitemShut {NoStop}%
\bibitem [{\citenamefont {Basov}\ \emph {et~al.}(1999)\citenamefont {Basov},
  \citenamefont {Woods}, \citenamefont {Katz}, \citenamefont {Singley},
  \citenamefont {Dynes}, \citenamefont {Xu}, \citenamefont {Hinks},
  \citenamefont {Homes},\ and\ \citenamefont {Strongin}}]{Basov:1999}%
  \BibitemOpen
  \bibfield  {author} {\bibinfo {author} {\bibfnamefont {D.~N.}\ \bibnamefont
  {Basov}}, \bibinfo {author} {\bibfnamefont {S.~I.}\ \bibnamefont {Woods}},
  \bibinfo {author} {\bibfnamefont {A.~S.}\ \bibnamefont {Katz}}, \bibinfo
  {author} {\bibfnamefont {E.~J.}\ \bibnamefont {Singley}}, \bibinfo {author}
  {\bibfnamefont {R.~C.}\ \bibnamefont {Dynes}}, \bibinfo {author}
  {\bibfnamefont {M.}~\bibnamefont {Xu}}, \bibinfo {author} {\bibfnamefont
  {D.~G.}\ \bibnamefont {Hinks}}, \bibinfo {author} {\bibfnamefont {C.~C.}\
  \bibnamefont {Homes}}, \ and\ \bibinfo {author} {\bibfnamefont
  {M.}~\bibnamefont {Strongin}},\ }\href {\doibase 10.1126/science.283.5398.49}
  {\bibfield  {journal} {\bibinfo  {journal} {Science}\ }\textbf {\bibinfo
  {volume} {283}},\ \bibinfo {pages} {49} (\bibinfo {year} {1999})}\BibitemShut
  {NoStop}%
\bibitem [{\citenamefont {Kirtley}\ \emph {et~al.}(1998)\citenamefont
  {Kirtley}, \citenamefont {Moler}, \citenamefont {Villard},\ and\
  \citenamefont {Maignan}}]{Kirtley:1998}%
  \BibitemOpen
  \bibfield  {author} {\bibinfo {author} {\bibfnamefont {J.~R.}\ \bibnamefont
  {Kirtley}}, \bibinfo {author} {\bibfnamefont {K.~A.}\ \bibnamefont {Moler}},
  \bibinfo {author} {\bibfnamefont {G.}~\bibnamefont {Villard}}, \ and\
  \bibinfo {author} {\bibfnamefont {A.}~\bibnamefont {Maignan}},\ }\href
  {\doibase 10.1103/PhysRevLett.81.2140} {\bibfield  {journal} {\bibinfo
  {journal} {Phys. Rev. Lett.}\ }\textbf {\bibinfo {volume} {81}},\ \bibinfo
  {pages} {2140} (\bibinfo {year} {1998})}\BibitemShut {NoStop}%
\bibitem [{\citenamefont {Loram}\ \emph {et~al.}(1994)\citenamefont {Loram},
  \citenamefont {Mirza}, \citenamefont {Wade}, \citenamefont {Cooper},\ and\
  \citenamefont {Liang}}]{Loram:1994}%
  \BibitemOpen
  \bibfield  {author} {\bibinfo {author} {\bibfnamefont {J.}~\bibnamefont
  {Loram}}, \bibinfo {author} {\bibfnamefont {K.}~\bibnamefont {Mirza}},
  \bibinfo {author} {\bibfnamefont {J.}~\bibnamefont {Wade}}, \bibinfo {author}
  {\bibfnamefont {J.}~\bibnamefont {Cooper}}, \ and\ \bibinfo {author}
  {\bibfnamefont {W.}~\bibnamefont {Liang}},\ }\href {\doibase
  https://doi.org/10.1016/0921-4534(94)91331-5} {\bibfield  {journal} {\bibinfo
   {journal} {Physica C: Superconductivity}\ }\textbf {\bibinfo {volume}
  {235-240}},\ \bibinfo {pages} {134 } (\bibinfo {year} {1994})}\BibitemShut
  {NoStop}%
\bibitem [{\citenamefont {Uchida}\ \emph {et~al.}(1996)\citenamefont {Uchida},
  \citenamefont {Tamasaku},\ and\ \citenamefont {Tajima}}]{Uchida:1996}%
  \BibitemOpen
  \bibfield  {author} {\bibinfo {author} {\bibfnamefont {S.}~\bibnamefont
  {Uchida}}, \bibinfo {author} {\bibfnamefont {K.}~\bibnamefont {Tamasaku}}, \
  and\ \bibinfo {author} {\bibfnamefont {S.}~\bibnamefont {Tajima}},\ }\href
  {\doibase 10.1103/PhysRevB.53.14558} {\bibfield  {journal} {\bibinfo
  {journal} {Phys. Rev. B}\ }\textbf {\bibinfo {volume} {53}},\ \bibinfo
  {pages} {14558} (\bibinfo {year} {1996})}\BibitemShut {NoStop}%
\bibitem [{\citenamefont {Nakamura}\ and\ \citenamefont
  {Uchida}(1993)}]{Nakamura:1993}%
  \BibitemOpen
  \bibfield  {author} {\bibinfo {author} {\bibfnamefont {Y.}~\bibnamefont
  {Nakamura}}\ and\ \bibinfo {author} {\bibfnamefont {S.}~\bibnamefont
  {Uchida}},\ }\href {\doibase 10.1103/PhysRevB.47.8369} {\bibfield  {journal}
  {\bibinfo  {journal} {Phys. Rev. B}\ }\textbf {\bibinfo {volume} {47}},\
  \bibinfo {pages} {8369} (\bibinfo {year} {1993})}\BibitemShut {NoStop}%
\bibitem [{\citenamefont {Basov}\ \emph {et~al.}(1994)\citenamefont {Basov},
  \citenamefont {Timusk}, \citenamefont {Dabrowski},\ and\ \citenamefont
  {Jorgensen}}]{Basov:1994}%
  \BibitemOpen
  \bibfield  {author} {\bibinfo {author} {\bibfnamefont {D.~N.}\ \bibnamefont
  {Basov}}, \bibinfo {author} {\bibfnamefont {T.}~\bibnamefont {Timusk}},
  \bibinfo {author} {\bibfnamefont {B.}~\bibnamefont {Dabrowski}}, \ and\
  \bibinfo {author} {\bibfnamefont {J.~D.}\ \bibnamefont {Jorgensen}},\ }\href
  {\doibase 10.1103/PhysRevB.50.3511} {\bibfield  {journal} {\bibinfo
  {journal} {Phys. Rev. B}\ }\textbf {\bibinfo {volume} {50}},\ \bibinfo
  {pages} {3511} (\bibinfo {year} {1994})}\BibitemShut {NoStop}%
\bibitem [{\citenamefont {Homes}\ \emph {et~al.}(2004)\citenamefont {Homes},
  \citenamefont {Dordevic}, \citenamefont {Strongin}, \citenamefont {Bonn},
  \citenamefont {Liang}, \citenamefont {Hardy}, \citenamefont {Komiya},
  \citenamefont {Ando}, \citenamefont {Yu}, \citenamefont {Kaneko},
  \citenamefont {Zhao}, \citenamefont {Greven}, \citenamefont {Basov},\ and\
  \citenamefont {Timusk}}]{Homes:2004}%
  \BibitemOpen
  \bibfield  {author} {\bibinfo {author} {\bibfnamefont {C.~C.}\ \bibnamefont
  {Homes}}, \bibinfo {author} {\bibfnamefont {S.~V.}\ \bibnamefont {Dordevic}},
  \bibinfo {author} {\bibfnamefont {M.}~\bibnamefont {Strongin}}, \bibinfo
  {author} {\bibfnamefont {D.~A.}\ \bibnamefont {Bonn}}, \bibinfo {author}
  {\bibfnamefont {R.}~\bibnamefont {Liang}}, \bibinfo {author} {\bibfnamefont
  {W.~N.}\ \bibnamefont {Hardy}}, \bibinfo {author} {\bibfnamefont
  {S.}~\bibnamefont {Komiya}}, \bibinfo {author} {\bibfnamefont
  {Y.}~\bibnamefont {Ando}}, \bibinfo {author} {\bibfnamefont {G.}~\bibnamefont
  {Yu}}, \bibinfo {author} {\bibfnamefont {N.}~\bibnamefont {Kaneko}}, \bibinfo
  {author} {\bibfnamefont {X.}~\bibnamefont {Zhao}}, \bibinfo {author}
  {\bibfnamefont {M.}~\bibnamefont {Greven}}, \bibinfo {author} {\bibfnamefont
  {D.~N.}\ \bibnamefont {Basov}}, \ and\ \bibinfo {author} {\bibfnamefont
  {T.}~\bibnamefont {Timusk}},\ }\href {\doibase 10.1038/nature02673}
  {\bibfield  {journal} {\bibinfo  {journal} {Nature}\ }\textbf {\bibinfo
  {volume} {430}},\ \bibinfo {pages} {539} (\bibinfo {year}
  {2004})}\BibitemShut {NoStop}%
\bibitem [{\citenamefont {Chakravarty}(1996)}]{Chakravarty:1996}%
  \BibitemOpen
  \bibfield  {author} {\bibinfo {author} {\bibfnamefont {S.}~\bibnamefont
  {Chakravarty}},\ }\href {\doibase 10.1103/PhysRevLett.77.4446} {\bibfield
  {journal} {\bibinfo  {journal} {Phys. Rev. Lett.}\ }\textbf {\bibinfo
  {volume} {77}},\ \bibinfo {pages} {4446} (\bibinfo {year}
  {1996})}\BibitemShut {NoStop}%
\bibitem [{\citenamefont {Sylju\aa{}sen}\ \emph {et~al.}(1997)\citenamefont
  {Sylju\aa{}sen}, \citenamefont {Chakravarty},\ and\ \citenamefont
  {Greven}}]{Chakravarty:1997}%
  \BibitemOpen
  \bibfield  {author} {\bibinfo {author} {\bibfnamefont {O.~F.}\ \bibnamefont
  {Sylju\aa{}sen}}, \bibinfo {author} {\bibfnamefont {S.}~\bibnamefont
  {Chakravarty}}, \ and\ \bibinfo {author} {\bibfnamefont {M.}~\bibnamefont
  {Greven}},\ }\href {\doibase 10.1103/PhysRevLett.78.4115} {\bibfield
  {journal} {\bibinfo  {journal} {Phys. Rev. Lett.}\ }\textbf {\bibinfo
  {volume} {78}},\ \bibinfo {pages} {4115} (\bibinfo {year}
  {1997})}\BibitemShut {NoStop}%
\bibitem [{\citenamefont {Kopp}\ \emph {et~al.}(2007)\citenamefont {Kopp},
  \citenamefont {Ghosal},\ and\ \citenamefont {Chakravarty}}]{Kopp:2007}%
  \BibitemOpen
  \bibfield  {author} {\bibinfo {author} {\bibfnamefont {A.}~\bibnamefont
  {Kopp}}, \bibinfo {author} {\bibfnamefont {A.}~\bibnamefont {Ghosal}}, \ and\
  \bibinfo {author} {\bibfnamefont {S.}~\bibnamefont {Chakravarty}},\
  }\href@noop {} {\bibfield  {journal} {\bibinfo  {journal} {Proc. National
  Acad. Sciences, USA}\ }\textbf {\bibinfo {volume} {104}},\ \bibinfo {pages}
  {6123} (\bibinfo {year} {2007})}\BibitemShut {NoStop}%
\bibitem [{\citenamefont {Koshi}\ \emph {et~al.}(2018)\citenamefont {Koshi},
  \citenamefont {Tadashi}, \citenamefont {Kensuke}, \citenamefont {Yasushi},
  \citenamefont {Takayuki}, \citenamefont {Takashi}, \citenamefont {Hitoshi},
  \citenamefont {Isao.W.}, \citenamefont {Masanori}, \citenamefont {Akihiro},
  \citenamefont {Ryosuke},\ and\ \citenamefont {Koike}}]{Kurashima:2018}%
  \BibitemOpen
  \bibfield  {author} {\bibinfo {author} {\bibfnamefont {K.}~\bibnamefont
  {Koshi}}, \bibinfo {author} {\bibfnamefont {A.}~\bibnamefont {Tadashi}},
  \bibinfo {author} {\bibfnamefont {M.~S.}\ \bibnamefont {Kensuke}}, \bibinfo
  {author} {\bibfnamefont {F.}~\bibnamefont {Yasushi}}, \bibinfo {author}
  {\bibfnamefont {K.}~\bibnamefont {Takayuki}}, \bibinfo {author}
  {\bibfnamefont {N.}~\bibnamefont {Takashi}}, \bibinfo {author} {\bibfnamefont
  {M.}~\bibnamefont {Hitoshi}}, \bibinfo {author} {\bibnamefont {Isao.W.}},
  \bibinfo {author} {\bibfnamefont {M.}~\bibnamefont {Masanori}}, \bibinfo
  {author} {\bibfnamefont {K.}~\bibnamefont {Akihiro}}, \bibinfo {author}
  {\bibfnamefont {K.}~\bibnamefont {Ryosuke}}, \ and\ \bibinfo {author}
  {\bibfnamefont {Y.}~\bibnamefont {Koike}},\ }\href@noop {} {\bibfield
  {journal} {\bibinfo  {journal} {Phys. Rev. Lett,}\ }\textbf {\bibinfo
  {volume} {121}},\ \bibinfo {pages} {057002} (\bibinfo {year}
  {2018})}\BibitemShut {NoStop}%
\bibitem [{\citenamefont {Sarkar}\ \emph {et~al.}(2020)\citenamefont {Sarkar},
  \citenamefont {Wei}, \citenamefont {Zhang}, \citenamefont {Poniatowski},
  \citenamefont {Mandal}, \citenamefont {Kapitulnik},\ and\ \citenamefont
  {Greene}}]{Sarkar:2020}%
  \BibitemOpen
  \bibfield  {author} {\bibinfo {author} {\bibfnamefont {T.}~\bibnamefont
  {Sarkar}}, \bibinfo {author} {\bibfnamefont {D.~S.}\ \bibnamefont {Wei}},
  \bibinfo {author} {\bibfnamefont {J.}~\bibnamefont {Zhang}}, \bibinfo
  {author} {\bibfnamefont {N.~R.}\ \bibnamefont {Poniatowski}}, \bibinfo
  {author} {\bibfnamefont {P.~R.}\ \bibnamefont {Mandal}}, \bibinfo {author}
  {\bibfnamefont {A.}~\bibnamefont {Kapitulnik}}, \ and\ \bibinfo {author}
  {\bibfnamefont {R.~L.}\ \bibnamefont {Greene}},\ }\href@noop {} {\bibfield
  {journal} {\bibinfo  {journal} {Science}\ }\textbf {\bibinfo {volume}
  {368}},\ \bibinfo {pages} {532} (\bibinfo {year} {2020})}\BibitemShut
  {NoStop}%
\end{thebibliography}
%

\end{document}